\documentclass[11pt]{article}
\usepackage{amsmath,amssymb,amsfonts}
\usepackage{caption}
\usepackage[font={scriptsize}]{subcaption}
\usepackage{setspace}
\usepackage[colorlinks=true, allcolors=blue]{hyperref}
\newcommand{\be}{\begin{equation}}
\newcommand{\ee}{\end{equation}}
\newcommand{\bea}{\begin{eqnarray}}
\newcommand{\eea}{\end{eqnarray}}
\newcommand{\ba}{\begin{array}}
\newcommand{\ea}{\end{array}}

\newcommand{\N}{\mathcal{N}}
\newcommand{\D}{\mathcal{D}}

\newcommand{\que}{\mathord{?}}

\newcommand{\RNum}[1]{\uppercase\expandafter{\romannumeral #1\relax}}

\begin{document}

\begin{titlepage}
\begin{center}
{\hbox to\hsize{hep-ph/2312.09166}}

\bigskip
\vspace{3\baselineskip}

{\Large \bf The Hierarchy Problem and $\N=2~ \D=5$ Supergravity} 

\bigskip

\bigskip

{\bf  Safinaz Salem}\\
\smallskip

{ \small \it  
 Department of Physics, Faculty of Science, Al Azhar University, Cairo 11765, Egypt\\

and

Center for Fundamental Physics, Zewail City of Science
and Technology, 6th of October City, Giza 12578, Egypt}

\bigskip

{\tt  safinaz.salem@azhar.edu.eg}

\bigskip

\vspace*{.5cm}

{\bf Abstract}\\
\end{center}
\noindent

We introduce a new scenario to solve the hierarchy problem based on $\N=2$, five-dimensional supergravity compactified on Calabi-Yau threefold down from $\D=11$ supergravity. When modeling the universe as a 3-brane embedded in a five-dimensional bulk, the background metric is proportional to one of the hypermultiplets fields, namely the dilaton, the volume modulus of the Calabi-Yau space. When solving
the dilaton field equation, we find that it is dependent on the extra dimension. We solve the modified Friedmann equations and find the implications of the model on the relation between the effective four-dimensional gravity scale in the brane and the Planck scale of the fifth extra dimension ($M$). 
A couple of different scales are considered for the extra dimension, the first where the extra dimension is in range of the solar system and 
$M$ is of order of the electroweak scale, and the second where the scale of the extra dimension is of order 
$\mu m$ and the higher dimensional Planck scale $M \sim 10^{13} ~ \text{GeV}$.
In both cases, the signatures of Kaluza-Klein excitations are quite distinct from the signatures of any previous extra dimensions model.

\bigskip

\bigskip

\end{titlepage}


\section{Introduction}

Why is the gravitational force in our universe so weak compared to the other forces of nature$\que$ 
The answer to this question relies on the so-called hierarchy problem, where the scale at which gravity unifies with the other natural forces is as high as $M_{pl} \sim 10^{19}~ \text{GeV}$, while the electroweak unification happens at only around $M_W = (100 -1000)~ \text{GeV}$

There have been many studies that tried to resolve the hierarchy between these two fundamental scales, the Planck scale, and the electroweak scale, whether within the context of the string theory \cite{Giddings:2001yu, Antoniadis:1998ig}, or within wrapped compactified extra dimensions like Randall-Sundrum model type \RNum{1} \cite{ Randall:1999ee}, or within infinite extra dimensions 
\cite{Gogberashvili:1999tb,Randall:1999vf}.

In type \RNum{1} string theory the relation between the Planck scale and transverse extra dimensions is given by \cite{Antoniadis2015, Antoniadis:1999cf} : 
\be 
M_p \sim M_s^{2+n} R^n_\perp,
\ee
 $R_\perp$ is the radius of the compact transverse extra dimensions n. That means
 $M_W/M_P$ ratio can be enhanced at large extra dimensions where only gravity propagates.

The generalization of that scenario is the so-called ADD 
(Arkani-Hamed-Dimopoulos-Dvali) model \cite{Arkani-Hamed:1998jmv}, which states that for 4+n higher spacetime dimensions, the effective four-dimensional Planck scale is given by,
\be 
M_p = M^{n+2} V_n,
\ee
$V_n$ is the volume of the compact space which is considered flat in ADD model. Whereas, in the Randall-Sundrum model RS-\RNum{1} 
the extra dimension is wrapped due to the existence of an exponential factor.
In RS-\RNum{1} the standard model (SM) particles are confined to a 3-brane called the TeV brane which is at a distance 
$y=L$ on an orbifold from another hidden 3-brane called the Planck brane. The space between the two branes is an Anti-de Sitter (AdS) extra dimension where gravity propagates. 
On the seen brane, the hierarchy between the weak and the gravity scales is naturally generated due to an exponential factor proportional to the size of the extra dimension. While the case where $L$ is taken to be infinite and only a single brane remains at $y=0$ is known as
RS type-\RNum{2} model.

However, it is worth mentioning that there are vital issues concerning extra dimensions brane-worlds models, because in these models the Hubble parameter is proportional to the energy density on the brane $\rho$ rather than being proportional to $\sqrt{\rho}$ as in the standard cosmology. That results in a non-convetional cosmic evolution of the brane-universe not like our universe \cite{Langlois:2004kc,Binetruy:1999ut}.
In this paper, we aim to introduce a new approach to solve the hierarchy problem based on modeling
the universe as a 3-brane embedded in the bulk of an ungauged five-dimensional $\D=5$ $\N=2$ supergravity 
(SUGRA$5$) that is compactified on Calabi-Yau threefolds from $\D=11$ supergravity. This model
has been studied before in many references like \cite{Gauntlett:2002nw, Bergshoeff:2004kh, Gutperle:2000sb}, while the model's cosmological implications
have been investigated in \cite{Emam:2020oyb, Emam:2023idq} where it has been shown that the cosmic evolution of the brane's scale factor coincides with that of our universe. Also, the various cosmological eras of the universe from the inflation until the late-time acceleration 
can be successfully interpreted through SUGRA$5$ due to the effects of the bulk and the topology of the theory.
So that the dark energy that causes the recent rapid accelerated expansion of the universe \cite{SupernovaSearchTeam:1998fmf}, namely the cosmological hierarchy \cite{Carroll:1992} can be explained. Consequently, our purpose here is to address the electroweak hierarchy.   

Utterly to the static solutions of RS models, we consider a time-dependent scale factor for the brane-universe. 
The bulk is filled by the hypermultiplets of SUGRA5, and we assume that the brane is filled by a perfect fluid.
While RS-\RNum{1} model generates a natural exponential hierarchy between the gravity and the weak scales on one 
of the model's two branes. In this model, there is a single brane embedded in a finite de Sitter extra dimension. We will show that
the effective $4\D$ Planck scale $M_{pl}$ is directly related to the five-dimensional Planck scale $M$ by the extra dimension size,
hence, $M_{pl}$ can be suppressed to the electroweak scale $M_{EW}= 1000 ~\text{GeV}$ or $\sim 10^{13}~ \text{GeV}$ at relevant extra dimension scales. The solution of the modified Einstein field equations is Bogomol'nyi-Prasad-Sommerfield (BPS) which implicates conserving supersymmetry (SUSY) only in the bulk \cite{Celi:2003qk, Ceresole:2000jd}.  

The paper is arranged as follows: In section (\ref{D=5SUGRA}) we revise the action of the SUGRA5 model in a symplectic form, and we
introduce the metric of the five-dimensional spacetime and the matter content of the theory. In section (\ref{hyper-scalars}) we solve the field equations of the hyper-scalars. In section (\ref{solve-EFE}) we solve the modified Friedmann equations, then we explore the model's implications on the relation between effective $4\D$ gravity scale $M_{pl}$ and the weak scale. We show that the electroweak hierarchy can be enhanced due to the existence of an extra dimension. We find the $4\D$ gravity limit of the theory and show how the cosmological hierarchy can be implicitly solved trivially according to the solution. At last, we present the final solution of the theory.

\section{$\D=5 ~\N=2$ supergravity and the spacetime metric}\label{D=5SUGRA}

The dimensional reduction of $D=11$ supergravity theory \cite{Emam:2010kt} over a Calabi-Yau 3-folds yields an ungauged $\N=2$ supersymmetric gravity theory in $D=5$ with a matter sector comprised of four scalar fields and their superpartners; collectively known as the universal hypermultiplet (UH). These are: the dilaton $\sigma$, which is proportional to the natural logarithm of the volume of the Calabi-Yau submanifold, the universal axion $\varphi$, the pseudo-scalar axion $\chi$ and its complex conjugate $\bar \chi$.
The action of the theory is the sum of the gravitational action and the bosonic action:
\bea%
\nonumber \mathcal{S}&=& \mathcal{S}_{\text{gravity}} + \mathcal{S}_{{matter}} \nonumber \\
\mathcal{S}_{\text{gravity}} &=& \int d x^4 ~\int dy~ \sqrt{-G} ~ \left (M^3 R \right) \nonumber \\
\mathcal{S}_{\text{matter}} &=&   \int_5 \left[  - \frac{1}{2}d\sigma  \wedge \star d\sigma  - e^\sigma  d\chi  \wedge \star d\bar \chi  - \frac{1}{2} e^{2\sigma } \left( d\varphi + \frac{j}{2}f \right) \wedge \star \left( d\varphi - \frac{j}{2}\bar f \right) \right],
\label{Action}
\eea
The action is in the symplectic representation \cite{Emam:2013, Canestaro:2013xsa}, where $\star$ is the $\D=5$ Hodge duality operator, $ f = \left( {\chi d\bar \chi  - \bar \chi d\chi } \right)$,
 and  $\bar f =  - f.$
For the spacetime background, we consider a five-dimensional spacetime metric of the form \cite{Binetruy:1999hy,Kim:2003pc}: 
\be 
ds^2 = g_{MN} ~ dx^M dx^N = e^{2 A\sigma (y)} ~ \eta_{\mu\nu} ~ dx^\mu dx^\nu + e^{2 B\sigma (y)} dy^2,
\ee
where $y$ is the transverse coordinate of the wrapped extra dimension. The hypersurface located at $y=0$ identifies the brane that forms our universe. $ e^{2 A \sigma (y)}$ is a wrap factor, and the factor $ e^{2 B \sigma (y)}$ to keep the metric conformally flat, $A$ and $B$ are constants, and $\sigma$ is the dilaton field. Contrary to the static RS solution, we seek a cosmological solution, so we choose the metric of the form:
\be
ds^2 = e^{2 A \sigma (y)} ~ [ - dt^2 + a^2 (t)~ d\bar{x}^2 ] +  e^{2 B\sigma (y)}dy^2,
\label{metric}
\ee
which preserves the homogeneity and isotropy of space since the brane's scale 
factor $(a)$ depends on time, and which induces an effective $\D=4$ FWR metric on the 3-brane.
The energy-momentum tensor can be split into two pieces
\be
T_{MN} = T_{MN}^{bulk} + T_{MN}^{brane},
\ee
where $\left(M, N = \mu, \nu , y\right)$ , $\left(\mu, \nu =0, 1,2, 3\right)$ , and $T_{MN}^{bulk} = \frac{-2}{\sqrt{-g}} \frac{\delta S_M}{\delta g^{MN}}$ is the bulk's energy-momentum tensor, which consists of the UH fields as following:
\bea
 T^{{\rm bulk}}_{\mu \nu}  &=&  \frac{1}{4}g _{\mu \nu} \left( {\partial _y  \sigma } \right)\left( {\partial ^y  \sigma } \right)  +\frac{1}{2}g _{\mu \nu} e^\sigma  \left( {\partial _y  \chi } \right)\left( {\partial ^y  \bar \chi } \right)
 +  \frac{1}{4}g _{\mu \nu} e^{2\sigma } \left( {\partial _y \varphi + \frac{j}{2}f_y } \right)\left( {\partial ^y \varphi - \frac{j}{2}\bar f^y } \right)  \nonumber\\
 T^{{\rm bulk}}_{yy}  &=&  \frac{1}{4}g _{yy } \left( {\partial _y  \sigma } \right)\left( {\partial ^y  \sigma } \right) -\frac{1}{2}\left( {\partial _y  \sigma } \right)\left( {\partial _y  \sigma } \right)
  +    \frac{1}{2}e^{\sigma } g _{y y }   \left( {\partial _y  \chi } \right)\left( {\partial ^y  \bar \chi } \right)-e^\sigma  \left( {\partial _y  \chi } \right)\left( {\partial _y  \bar \chi } \right) \nonumber\\
  &+&   \frac{1}{4}e^{2\sigma } g _{y y } \left( {\partial _y \varphi + \frac{j}{2}f_y } \right)\left( {\partial ^y \varphi - \frac{j}{2}\bar f^y } \right) - \frac{1}{2}e^{2\sigma } \left( {\partial _y \varphi + \frac{j}{2}f_y } \right)\left( {\partial _y \varphi - \frac{j}{2}\bar f_y } \right),\label{Bulk Stress tensor}
\eea
Whereas $T_{MN}^{brane}$ represents the matter content in the brane, which we assume it is
a perfect fluid, so that:
\be
T_{MN}^{brane} = e^{2 A \sigma} \text{diag} ~ ( \rho , a^2 P, a^2 P, a^2 P , 0),
\ee
The energy density $\rho$ and the pressure $P$ are independent of the position inside the brane. 
The bulk's stress tensor can be simplified by considering the BPS condition \cite{Canestaro:2013xsa}:
\be
    d\sigma  \wedge \star d\sigma  -  e^{2\sigma } \left( {d\varphi + \frac{j}{2}f} \right) \wedge \star\left( {d\varphi - \frac{j}{2}\bar f} \right) + 2e^\sigma  d\chi  \wedge \star d\bar \chi  = 0.\label{BPS condition}
\ee
which implies partially conserving SUSY in the theory by preserving it only in the bulk. That imposing the vanishing of the supersymmetric transformations of the superpartners. Accordingly, the components of the total stress tensor are given by:                              
\bea
    T_{00}  &=&  - \frac{1}{2} \sigma '^2  + \rho e^{2 A \sigma } \nonumber\\
    T_{11}  &=& \frac{a^2}{2} \sigma '^2 + p a^2 e^{2 A \sigma }  \nonumber\\
    T_{yy}  &=&  - \frac{1}{2}\sigma '^2  \nonumber\\
    T_{0y}  &=& 0,
\eea
where a prime is a derivative with respect to $y$. Being $T_{0y} =T^{{\rm brane}}_{yy}= 0$ guarantees the confinement of the standard model particles to the brane, i.e. they are not allowed to diffuse to the bulk.
\section{The solutions of the hyper-scalars field equations }\label{hyper-scalars}

To depict the dilaton behavior with respect to the bulk, we solve the field equation of the hyper-scalars. The field equation can be derived from the variation of the action (\ref{Action}). The field equations for $\sigma$, $\left(\chi, \bar\chi\right)$ and $\varphi$ respectively are \cite{Emam:2020oyb}:
\bea
    \left( {\Delta \sigma } \right)\star \mathbf{1} - e^\sigma  d\chi  \wedge \star d\bar \chi  - e^{2\sigma } \left( {d\varphi + \frac{j}{2}f} \right) \wedge \star\left( { d\varphi - \frac{j}{2}\bar f} \right) &=& 0 \label{Dilaton eomSPLIT}\\
    d^{\dagger} \left[ {e^\sigma  d\chi  + je^{2\sigma } \chi \left( {d\varphi + \frac{j}{2}f} \right)} \right] &=& 0 \label{Chi eomSPLIT}\\
    d^{\dagger} \left[ {e^\sigma  d\bar \chi  - je^{2\sigma } \bar \chi \left( {d\varphi - \frac{j}{2}\bar f} \right)} \right] &=& 0
    \label{Chi bar eomSPLIT}\\
    d^{\dagger} \left[ {e^{2\sigma } \left( {d\varphi + \frac{j}{2}f} \right)} \right] &=& 0\label{a eomSPLIT},
\eea
where $d^\dag$ is the adjoint exterior derivative and $\Delta$ is the Laplace-De Rahm operator. 
Integrating (\ref{a eomSPLIT}) once yields: 
\be
    e^{2\sigma } \left( {d\varphi + \frac{j}{2}f} \right) =  dh, \label{hequ}
\ee
where $h\left(y\right)$ is a harmonic function; $\Delta h = d^\dagger dh = 0$ and we take the integration constant equals unity.
Using the BPS condition (\ref{BPS condition}) and equation (\ref{hequ}) into the dilaton equation (\ref{Dilaton eomSPLIT}) gives
\be
    \left( {\Delta \sigma } \right)\star \mathbf{1} + \frac{1}{2}d\sigma  \wedge \star d\sigma  =  { \frac{{3 }}{2}}  e^{ - 2\sigma } d h \wedge \star d h.\label{Dilaton a BPS}
\ee
or in a non-symplectic form;
\be
\sigma'' + \frac{1}{2} \sigma'^2 = \frac{3}{2} e^{-2\sigma} h'^2.
\label{dilaton}
\ee
$\nabla^2 h=0$ yeilds $h''=0$ with a solution
\be
h(y)=c_1+y c_2,
\label{Harmonic}
\ee
from the initial conditions the integration constants are
$c_1=0$ and $c_2=1$. So the dilaton equation (\ref{dilaton}) becomes: 
\be
\sigma'' + \frac{1}{2} \sigma'^2 -\frac{3}{2} e^{-2\sigma} =0.
\ee
When solving it gives:
\be
\sigma =  Log [1 + C~ y^2],
\label{sigma1}
\ee
where $C=0.75$. The axion equations (\ref{Chi eomSPLIT}, \ref{Chi bar eomSPLIT}) can be integrated once to give
\bea
    d\chi  + nj\chi e^{ - \sigma } d h = 0 \nonumber\\
    d\bar \chi  - nj\bar \chi e^{ - \sigma } d h = 0,\label{TheCHIs}
\eea
Using equations (\ref{Harmonic}) and (\ref{sigma1}) in these equations, yields:
\bea
    \frac{d\chi}{\chi}  = - j (1+ C y^2)^{-1} d y  \nonumber\\
    \frac{d\bar \chi}{\bar \chi }  = j(1+ C y^2)^{-1} d y ,
\eea
When integrating it gives: 
\bea
 \chi = A \left( \frac{C y+i}{C y-i} \right)^4 \nonumber\\
 \bar \chi = \bar{A} \left( \frac{C y-i}{C y+i} \right)^4  ,
\eea
where $A$ is a split-complex constant of integration having the property of being null ( $\left| A \right|^2  = A\bar A = 0$). 
This makes $\chi$ null as well. From which the function $f$ in (\ref{Action}) is vanished. 
The equation of the universal axion (\ref{hequ}) becomes:
\be
  d  \varphi =  (1+C y^2)^{-2} dy, 
\ee
integrating:
\be
\varphi = \frac{y}{2 \left(C y^2+1\right)}+\frac{\tan ^{-1}\left(\sqrt{C} y\right)}{2 \sqrt{C}}+ c_3.
\ee
\section{Solving the modified Friedmann equations}\label{solve-EFE}
Following the metric (\ref{metric}), the non-vanishing components of the Einstein tensor $G_{MN}$ are:
\bea%
G_{00}&=& 3 \left( \frac{\dot a}{a} \right)^2 + 3 A ~e^{2(A-B) \sigma} ~  \left[ \left(2A-B\right) \sigma'^2+\sigma'' \right], \nonumber \\
G_{11}&=&- \left( 2{\ddot a}a + \dot a^2  \right)   +3 A a^2 ~e^{2(A-B) \sigma} ~ \left[ \left(2A-B\right) \sigma'^2+ \sigma'' \right],\\ 
G_{yy}&=& 6 A^2 \sigma'^2 -3 e^{2(B-A)\sigma} \left[ \frac{\ddot a}{a} +  \left(\frac{\dot a}{a} \right)^2  \right].
\eea
Then the Einstein's equations $G_{MN} + g_{MN} \Lambda = \kappa ^2 T_{MN}$ yields:
\bea%
 3 H^2 &=& - 3 A ~e^{2(A-B) \sigma} ~  \left[ \left(2A-B\right) \sigma'^2+\sigma'' \right]  - \frac{\sigma'^2}{2}  + (\Lambda + \rho) e^{2 A \sigma} \nonumber \\
2 \left( \frac{\ddot a}{a}  \right) + H^2 &=&   3 A ~e^{2(A-B) \sigma} ~ \left[ \left(2A-B\right) \sigma'^2(y)+ \sigma'' \right] - \frac{\sigma'^2}{2} + ( \Lambda-P) e^{2A\sigma} \nonumber \\
3 e^{2(B-A)\sigma} \left[ \frac{\ddot a}{a} +  H^2  \right]&=&\frac{\sigma '^2}{2}+ 6 A \sigma '^2 + \tilde{\Lambda} e^{2B\sigma},\label{Eq}
\eea 
where $H=\left( \frac{\dot a}{a} \right) $ is the brane's Hubble parameter and $\tilde{\Lambda}$ is the cosmological constant of the bulk. 
Eliminating the brane's scale factor, we get  a single equation in $\sigma(y)$
\be
\sigma'^2 \left(13 e^{2 (A-B) \sigma}+2\right)+ \left(2 \tilde{\Lambda}- 4 \Lambda - \rho + 3 P \right) e^{2 A \sigma}=6 A e^{2 c \sigma} \left( (2 A-B) \sigma '^2+\sigma '' \right)
\ee
The fitting of this equation solution to the dilaton equation (\ref{sigma1}) implies that  
$A=-0.2$ and $B=3.4 $, while $\left(2 \tilde{\Lambda}- 4 \Lambda - \rho + 3 P \right)=-2$.   
To understand how the gravity behaves relative to the extra dimension, perturbate the gravity part of the action (\ref{Action})
around the background metric: 
\be
ds^2 = e^{-0.4~ \sigma (y)} ~ (\eta_{\mu\nu}+h_{\mu\nu} ) ~ dx^\mu dx^\nu + e^{6.8~ \sigma (y)} dy^2,
\label{metric2}
\ee
where $h_{\mu\nu}$ is the effective four-dimensional graviton field which appears as small perturbation around the theory metric (\ref{metric}). So that:
\be
\det (-G) = \det(-\bar{g}) e^{6.4\sigma},
\ee
and the four-dimensional effective action then follows:
\be 
\mathcal{S}_{eff} \supset M^3 \int d^4 x \int d y ~  e^{2\sigma}~ \sqrt{-\bar{g}} ~ \bar{R}, 
\ee
where $\bar{R}$ denotes the four-dimensional Ricci scalar made out of $\bar{g}_{\mu\nu}$, while $R$ is the five-dimensional Ricci scalar made out of $\bar{G}_{MN}$. Leading to the relation between the effective $4\D$ Planck scale $M_{pl}$ and $\D=5$ Planck scale $M$ :
\be
M^2_{pl} = M^3 ~ \int d y ~ e^{2\sigma}.
\ee
From equation (\ref{sigma1}), $e^{2\sigma} \propto y^2$, so $\int d y ~ e^{2\sigma} \sim \int d y ~  y^2 = \frac{1}{3} y^3+ n $ . Then the relation between the Planck scale of the four-dimensional effective theory $M_{pl}$ and the five-dimensional Planck scale then will be 
\be 
M^2_{pl} = M^3 ~ \left[ \frac{1}{3} ~y^3 + n \right]
\label{PandM}
\ee
Where $n$ is the constant of integration. If $n$ is taken to be zero, the bulk extra dimension can not be infinite, and that we consider here. Let us assume two different scenarios, the first where the Planck scale of the extra dimension is of order of the electroweak scale $M= 10^{3} ~ \text{GeV}$, which gives $y$ in order of the solar system ranges $y \sim 10^7 ~ \text{cm}$. Where to avoid contradicting the general relativity, the crossover scale $y^3$ should be around $10^8$ times bigger than the solar system size \cite{Dvali:2000hr}. 
The second scenario is to assume that the bulk extra dimension is  $( \mathcal{O}) ~1 ~ \mu \text{m}$ , which gives $M=  10^{13} ~ \text{GeV}$. This meets the constraints on the size of the large extra dimensions compared to the Planck length,  where gravity is tested if it deviates from the inverse square $\left(1/r^2\right)$ law. In \cite{Hoyle:2004cw} for instance, nothing new has been found till around $30~\mu$m. 

Now let us discuss the physical implications of each case. 
The first scenario is an alternative to compactification where many models have been proposed \cite{Gregory:2000jc, Fonseca:2010va, Ito:2002tx, Neupane:2010ey}. Accordingly, it is manifestation that the extra dimensions should not be small or finite and it depend on the gravitational potential and the crossover scale between $M_{pl}$ and $M$.

In the second scenario, fine-tuning requires to introduce a new parameter $M~ (\mathcal{O} )\sim 10^{13} ~ \text{GeV}$, which is better than $10^{19} ~ \text{GeV}$. According to this case, the extra dimension is wrapped like the compactified extra dimensions. At $y$ equals micrometers gravity or namely the graviton propagates while the SM particles are confined to the 3- brane. That causes the observed weakness of gravity in the brane-universe. However in this case we are facing like ADD model a new hierarchy between the compactification scale  $\mu= \left(1/y^3 \right) \sim 10^3 ~ \text{eV}$ and $M$. The coupling of each Kaluza-Klein (KK) excitation will be suppressed by $ \left(1/M \right) \sim 0.001 ~ eV^{-1}$, which is distinguished from KK signals in ADD model, whose couplings are of order $TeV^{-1}$, or KK signals in the RS model, whose couplings to matter alternatively $ \sim$ TeV. The existence of new light resonances may be dangerous because they can produce more CMB anisotropy and affect other astrophysical observations. Anyhow the coupling of KK excitations in our model is smaller than ADD's KK couplings, so the signatures of the KK tower according to our model are more reachable. While the restrictions from particles physics, astrophysics, and cosmology \cite{Mirabelli:1998rt,Fermi-LAT:2012zxd,CMS:2016crm, DAmbrosio:2017wis} are not valid.

The Hubble parameter can be found from the modified Einstein's equations (\ref{Eq}) by eliminating $F(\sigma) = (3 A ~e^{2(A-B) \sigma} ~  \left[ \left(2A-B\right) \sigma'^2+\sigma'' \right] ) $ from the first and the second equation, then solving the resultant equation with the third equation, we get:
\be
H^2 = e^{2 A \sigma} \left( \Lambda - \frac{1}{3}\tilde{\Lambda} + \frac{1}{2} (\rho - p) \right) - 
  \frac{1}{2} \sigma'^2 - \frac{1}{6} (1 + 12 A) e^{2 (A - B) \sigma} \sigma'^2, 
\label{Hub}
\ee
\be
H^2 = e^{-0.4 \sigma} \left( \Lambda - \frac{1}{3} \tilde{\Lambda} + \frac{1}{2} (\rho - p) \right) + 
  \left ( 0.23 e^{-7.2 \sigma} - \frac{1}{2} \right) \sigma'^2 
\label{Hub}
\ee
Let us not that the terms $\Lambda$, $\rho$, and $\tilde{\Lambda}$ are observed from within the brane-universe, 
which is located at  $y=0$ so that they are independent of $y$, in other words
$\Lambda e^{-0.4\sigma}$, $\tilde{\Lambda} e^{-0.4\sigma}$, and $\rho  e^{-0.4\sigma}$ is evaluated at $e^{-0.4~\sigma} \sim 1 $.
Retriveng the standard field equations requires that:
\be
  \left ( 0.23 ~ e^{-7.2 \sigma} - \frac{1}{2} \right) \sigma'^2 =0
\label{sigma0}
\ee
This also means that on the brane's boundary the dilaton appears as a constant. Now since the cosmological hierarchy relies on the exceeding of the theoretical cosmological constant over the observed dark energy by ( $\mathcal{O} \sim 120$), because the first is related 
to the vacuum energy density, we define a new cosmological constant $ \Lambda_{0}= \Lambda - \frac{1}{3} \tilde{\Lambda} $.
Where $\Lambda$ considered the cosmological constant related to the vacuum energy, and $\Lambda_{0}$ is the 
observed dark energy responsible for the current observed accelerated expansion of the universe, and 
$\tilde{\Lambda}$ makes the required balance between the theoretical and the observed values of the dark energy. 
So the cosmological hierarchy problem can be implicitly solved within this model. Also, this implies that the bulk is a de Sitter space.
Integrating the Hubble parameter, and dropping the index $0$, we get brane's scale factor:
\be 
a(t) = n~ exp \left(\sqrt{\frac{2 \Lambda+\rho -p }{2}}~t \right),    
\ee
where $n=a[0]$ is the integration constant. Consider the 3-brane is a de Sitter space with $\rho=-p$ and $\Lambda \sim \rho$, then
\be 
a(t) = n~ exp \left(\sqrt{\Lambda } ~t \right),    
\ee
which is similar for the scale factor of general relativity in case of dark energy dominance.
Finally, the complete solution of the theory is given by:
\be
ds^2 = (1+ 0.75~ y^2)^{-0.4} ~ \left[ - dt^2 +  n^2~ e^{2\sqrt{\Lambda}~t} d\bar{x}^2 \right] +  (1+ 0.75~ y^2)^{6.8} dy^2,
\ee

\section{Conclusion}
We introduce an alternative approach to solve the tension between the main fundamental scales in our universe the Planck scale and the electroweak scale. Wherein, till nowadays  there are no approved reasons why gravity decouples from the other forces in nature and becomes strong only at a high energy scale. According to $\N=2, ~ \D=5$ supergravity the universe is modeled as a 3-brane embedded in 
a five-dimensional bulk, where the standard model particles are confined to the brane-universe and only the graviton can propagate to the bulk where UH fields exist. The five-dimensional metric is proportional to one of the hyper-scalars fields, the dilaton.
By solving the dillaton field equation and the modified Einstein field equations, we show that this field depends on the extra dimension.
When perturbing the action around the background metric, the $4\D$ effective gravitation appears as the zero modes of the theory and the 
behavior of the effective gravity in the bulk can be deduced. We find that the $4\D$ Planck scale  is related to the five-dimensional Planck 
scale ($M$) directly through the extra dimension scale $y$.
Since we have no limits on the extra dimension, unless it can not be infinite, we assume a couple of scales for $y$, the first
where $M$ ($\mathcal{O}$) the electroweak scale, so that, gravity unifies with the other forces of nature at only $1$ TeV, which leads to a large extra dimension around the solar system scale. Consequently, in this case the fundamental scale of nature is the electroweak scale instead of $M_{pl}$, also it indicates that the extra dimensions are not necessary to be only comapctified.
We also assume that $y$ ($\mathcal{O}$) micrometers, which leads to $M= 10^{13}$ GeV, so the effective Planck scale can be reduced by order $10^6$. The solutions are BPS that preserve SUSY only in the bulk while it is broken in the brane because it contains a perfect fluid.
We find the $4\D$ gravity limit of the theory in case of dark energy dominance. This model has many merits as shown in previous studies that it is a viable cosmology model, even the solution of the Friedmann equations here beholds an answer for the cosmological hierarchy, where the existence of a positive bulk cosmological constant can explain why the observed dark energy is much less than the theoretical cosmological constant assumed in the Einstein's equations.    
Indeed any theory needs experimental consequences to be proved. On the other hand, there are many experimental constraints on the existence of possible extra dimensions. One of our goals from this work is to reach for experimental signatures for  $\D=5$ supergravity, like the KK excitations that are distinguished from any other extra dimensions model and maybe that is an avenue for future research. Also 
another possible upcoming research path is to deduce the non-intuitive topology of the Calabi-Yau threefolds, since the dilaton field is related to its volume.

\end{document}